\documentclass[11pt]{article}
\usepackage{amsfonts}
\usepackage{epic}
\newcommand{\be}{\begin{equation}\label}
\newcommand{\ee}{\end{equation}}
\newcommand{\bea}{\begin{eqnarray}\label}
\newcommand{\eea}{\end{eqnarray}}

\newcommand{\tr}{\tilde{\rho}}

\usepackage{graphicx}

\begin{document}
\vspace{8cm}

\begin{center}
{\bf {\Large Mean-field theory for Heisenberg zigzag ladder:
Ground state energy and spontaneous symmetry breaking }}
\end{center}
\vspace{4cm}
\begin{center}
{\bf  Vagharsh~V~Mkhitaryan\footnote{e-mail:{\sl
vgho@mail.yerphi.am}}$^{,~a}$~and~~Tigran~A.~Sedrakyan\footnote{e-mail:{\sl
tigrans@physics.utah.edu}}$^{,~b}$}

\vspace{1cm}

{\em $^a$ Yerevan Physics Institute, Alikhanian Br. str. 2, Yerevan 36,  Armenia\\
and The Abdus Salam ICTP, Strada Costiera 11, Trieste 34014, Italy\\
 $^b$ Department of Physics, University of Utah, Salt Lake
City, UT 84112}\\

\vspace{1.5 cm}
{\it \large Dedicated to the memory of Daniel Arnaudon}
\end{center}
\vspace{2cm}
\begin{center}
{\LARGE  Abstract}
\end{center}
The spin-1/2 zig-zag Heisenberg ladder ($J_1 - J_2$ model) is
considered. A new representation for the model is found and a
saddle point approximation over the spin-liquid order parameter
$\langle \vec \sigma _{n-1}(\vec \sigma_{n}\times \vec \sigma
_{n+1}) \rangle$ is performed. Corresponding effective action is
derived and analytically analyzed. We observe the presence of
phase transitions at values $J_2/J_1=0.231$ and $J_2/J_1=1/2$.

\vspace{3cm}

\newpage

\section{Introduction}
\indent

Unconventional spin-liquid phases in frustrated spin chains
attracted notable theoretical and experimental interest in recent
years. The natural question is whether the frustrations in
antiferromagnetic Heisenberg chains can stabilize the new phases
with exotic spin excitations observed in ladder systems. The model
used to analyze the effect of frustration in antiferromagnetic
spin chains is the so called spin-1/2 $J_1-J_2$ model with the
Hamiltonian \be{j1j2} H=J_1\sum_n[\vec\sigma _n\vec \sigma _{n+1}
-1] + J_2\sum_n[\vec\sigma _n\vec \sigma _{n+2} -1], \ee where
$\vec \sigma _n=2\vec S_n$ are Pauli matrices. The bosonization
analysis of this model was performed by Haldane \cite{Haldane},
and the phase diagram has been studied intensively by various
authors \cite{Haldane1, Eggert,ON,ChP, WA, Ners} (for a review see
also \cite{Lech}).

The interest in this model is not purely theoretical. There are
 inorganic compounds in nature, such as  $Cs_2CuCl_4$ \cite{Coldea},
$CuGeO_3$ \cite{Hase}, $LiV_2O_5$ \cite{Fujiwara}, or $SrCuO_2$
\cite{Matsuda}, which can be described by the spin-1/2 $J_1-J_2$
chain Hamiltonian Eq.~(\ref{j1j2}).

The investigation of the spin-1/2 $J_1-J_2$ model's phase diagram
(\ref{j1j2}) starts from the weak coupling limit when $J_2 <<
J_1$. Classically, in this limit, the $J_1-J_2$ chain has a long
range Neel order. The excitations are massless spin-waves,
frustrated by an irrelevant perturbation. At the values  $J_2 /
J_1 > 1/4$, the spins in the ground state are arranged in a canted
configuration, in which each spin makes a fixed angle
$\alpha=\arccos[- J_1 / 4J_2]$ with its predecessor. The classical
ground state of the model is doubly degenerate since the spin
configurations can turn clockwise and counterclockwise with the
same energy.

In quantum field theory, it is believed that at larger values,
$J_2/J_1 > 1/4$, a phase transition of the
Berezinskii-Kosterlitz-Thouless (BKT) type \cite{BKT} occurs,
which separates the gapless spin-1/2 Heisenberg phase from a fully
massive region. This phase is characterized by the two-fold
degenerate dimerized ground state and a spontaneous breaking of
the lattice translation symmetry. Frustration stabilizes this
gapful phase. The actual value for the ratio of the coupling
constants at the transition point was found numerically in
\cite{Haldane1,Eggert,ON,ChP, WA} to be slightly lower than
$J_2/J_1=0.241$ due to quantum fluctuations.

At the larger value (when $J_2=1/2 J_1>0$) the model coincides
with the Majumdar-Ghosh (MG) model \cite{MG,ShSuth} and is exactly
integrable \cite{Bethe, Takhtaj}. The existence of a mass gap in
the MG model has been shown rigorously in \cite{Affleck}. The
correlation function is found to be zero at distances larger than
the lattice spacing. The ground state of the phase at $J_2/J_1
>1/2$ is found to be a condensate of dimerized singlets of pairs
of neighbor spins, which is $Z_2$ degenerate. This spontaneous
breaking of the $Z_2$ discrete symmetry creates a kink type of
topological excitation, the tails of which end in the different
$Z_2$ vacua at $\pm \infty$. The spectrum of this excitation is
massive, and the gap is decreasing at $J_2/J_1 \rightarrow
\infty$, when the system becomes a pair of noninteracting spin -
1/2 Heisenberg chains.

In the present article, we develop an approach to analyze the
model on the matter of critical behavior, based on the idea that
the middle phase, $1/4 <J_2/J_1 <1/2$, can be characterized by the
spin-liquid order parameter
\begin{eqnarray}
\label{order} \varphi=\langle\vec \sigma _{n-1}(\vec \sigma
_{n}\times \vec \sigma _{n+1})\rangle
\end{eqnarray}
defined on the triangles of the zig-zag chain. Performing a
mean-field (saddle point) approximation, we reduce the model to an
extended Heisenberg chain with topological term, which appears to
be integrable. Then, by use of the technique of Thermodynamic
Bethe Ansatz, we calculate the effective action (formula
(\ref{free})) as a function of the spin-liquid order parameter
$\varphi$.
The analysis of the effective action shows that, at the points
$J_2/J_1=0.230971$ and $J_2/J_1=1/2$, we indeed have phase
transitions, as it was expected.

\section{New representation for the $J_1 - J_2$ model}
\indent

For an alternative treatment of the Hamiltonian (\ref{j1j2}), we
make use of the following identity for the square of the Hermitian
operator $\chi_{a b c}=\vec \sigma _{a}(\vec \sigma _{b}\times
\vec \sigma _{c})$ (scalar chirality operator)

\begin{eqnarray}
\label{identity}
 \left ( \vec \sigma _{a}(\vec \sigma _{b}\times
\vec \sigma _{c}) \right )^2=-2 \Bigl(  [\vec\sigma _a\vec \sigma
_{b} -1]+ [\vec\sigma _a\vec \sigma _{c} -1]+[\vec\sigma _b\vec
\sigma _{c} -1] \Bigr).
\end{eqnarray}
On the right hand side of this identity, one may recognize a sum
of three different spin exchange terms, marked by mutually
non-coinciding indices $a\neq b\neq c$. The form of
Eq.~(\ref{identity}) suggests a transformation for the $J_1-J_2$
Hamiltonian, a dual representation of which would contain the
square of the above mentioned scalar chirality operator. Thus,
using Eq.~(\ref{identity}), we map the expression (\ref{j1j2}) to

\be{tj1j2} {\tilde H}=\sum_n[\vec\sigma_n\vec \sigma _{n+1} -1]
-\frac {g}2\sum_n \chi _{n,n+1,n+2}^{~2}, \qquad g=\frac
{J_2}{J_1-2J_2}, \ee were we have rescaled the Hamiltonian $H$ by
the constant factor $(J_1-2J_2)$, i.e.,
\begin{eqnarray}
\label{factor}
 \tilde{H}=\frac{H}{J_1-2J_2}.
\end{eqnarray}
This will be our starting point. We would like to emphasize
however, that, to the best of our knowledge, the Heisenberg zigzag
ladder ($J_1-J_2$ model) has not been represented in this form in
the literature previously.

\section{Mean-field theory and its Bethe Ansatz solution}
\indent

Our further analysis is close in spirit to that of Affleck and
Marston, applied in Ref. \cite {AM} for the solution of the
Heisenberg-Hubbard model. It is based on the approximation of the
partition function of the given model by an exactly integrable
one. Namely, we introduce a scalar field, $\phi _n$, by a
Hubbard-Stratanovich transformation, which maps the Hamiltonian
${\tilde H}$ to ${\cal H}$, where

\be{HS}
 {\cal H}={\tilde H}+\frac {g}2\sum_n(\phi _n-\chi _{n,n+1,n+2})^2 = \sum_n[\vec\sigma _n\vec \sigma _{n+1} -1] -g\sum_n\phi _n\chi _{n,n+1,n+2}+
\frac {g}2\sum_n\phi ^2_n. \ee This map induces a constant factor
in the partition function \be{pf}Z=Tr \left (\exp-\beta {\tilde
H}\right )= \mbox{const}\, Tr \left ( \int \exp-\beta {\cal
H}(\{\phi _n\})\prod d\phi _n\right ), \ee which however is
irrelevant. The fact that the transformation (\ref{HS}) leaves the
dynamics of the model unchanged may be shown using the coherent
state path integral representation for the partition function,
where the functional integral over the field $\{\phi _n\}$ can be
exactly evaluated. Below, we will investigate the zero temperature
limit, $\beta \rightarrow \infty $, of the integral on the right
hand side of Eq.~(\ref{pf}) in the saddle point approximation. We
are going to analyze the mean-field theory corresponding to a
certain saddle point, which we believe gives the main
contribution. The question of the existence of any other saddle
points, however, will be left for future investigations. More
precisely, we consider the saddle point equation
\begin{eqnarray}
\label{saddle} \frac{\partial {\cal H}}{\partial \phi_n}=0.
\end{eqnarray}
 The solution of Eq.~(\ref{saddle}) with regard to the bosonic fields
 $\{\phi_n\}$ can be obtained by the substitution of Eq.~(\ref{HS}) into Eq.
 (\ref{saddle}). It has the form of a set of $N$ (the lattice size) coupled equations
 \be{sp}
\phi _n=\frac{Tr \left ( \chi _{n,n+1,n+2}\,e^{-\beta {\cal
H}(\{\phi _n\})}\right )} {Tr \left (e^{-\beta {\cal H}(\{\phi
_n\})}\right )}\equiv \langle \chi _{n,n+1,n+2}\rangle , \ee where
$n=1\dots N$, and we have cyclic boundary conditions. It would be
reasonable to restrict ourselves by some (quasi) translational
invariant, homogeneous saddle points. Therefore, we consider the
solution where the operators $\chi _{n,n+1,n+2}$, for all
$n=1\dots N$, have the same mean value. Then the set of coupled
equations (\ref{sp}) simplifies, and for all values of $n$
acquires the form
 \be{qtisp}\langle \chi _{n,n+1,n+2}\rangle=\varphi.
\ee Thus, in this way, the original problem reduces to the
eigenfunction problem for the mean-field Hamiltonian \be{sh} {\cal
H}_M=\sum_n[\vec\sigma _n\vec \sigma _{n+1} -1] -\alpha \sum_n
\vec \sigma _{n}(\vec \sigma _{n+1}\times \vec \sigma
_{n+2}),\quad \alpha =g\varphi. \ee This model appears to be
exactly solvable by means of Bethe Ansatz. This is because the
second term in expression (\ref{sh}) commutes with the first term,
which, in turn, is the Heisenberg Hamiltonian. Therefore, skipping
the demonstration of the standard technique of Algebraic Bethe
Ansatz (since it repeats the one for the $XXZ$ model
\cite{Takhtaj}), we present here only the solution. The
eigenvectors can be parameterized through the set of parameters
(rapidities), $\{x_i\}$, which satisfy the set of Bethe equations
\be{BAE} \left(\frac{x_j-i}{x_j+i}\right)^N=-\prod\limits_{k=1}^M
\frac{x_j-x_k-2i}{x_j-x_k+2i}. \ee The corresponding state has a
total spin projection $S_z=N-M$ and energy \be{selfenergy}
E(x_1,..,x_M)=-\sum\limits_{j=1}^M\left(1+2\alpha
\partial_{x_j}\right) \frac 8 {x^2_j+1}.
\ee

\newpage
\section{Free energy}
\indent

In the present section, we present our calculations of the ground
state energy of the $J_1-J_2$ model in the mean-field
approximation. Namely, we calculate the free energy in the model
(\ref{sh}). The model (\ref{sh}) has been studied in Ref.
\cite{tsvelik} for a fixed value of the parameter $\alpha$.
Here,
we present our exact analytical calculations of the free energy of
the mean-field model (\ref{sh}), and analyze, in detail, the free
energy as a function of the parameter $\alpha$. The calculation is
based on the method of Thermodynamic Bethe Ansatz, introduced in
\cite{Yangs} (for details see also \cite{Taka}). By definition,
the thermodynamic limit is given by the following conditions
\begin{eqnarray}
\label{termlim} N\rightarrow \infty,\quad M\rightarrow
\infty,\quad \frac{M}{N}= const.
\end{eqnarray}
In the thermodynamic limit (\ref{termlim}), the Bethe equations
(\ref{BAE}) become integral equations. In order to represent these
integral equations in a convenient form, we introduce the
densities $ \rho (t)$, defined as $ \rho
(t)=\Bigl(\frac{dx}{dt}\big | _{t=t(x)}\Bigr)^{-1} $. Then the
equations (\ref{BAE}) reduce to the following system of $N$
integral equations for the densities

\be{dis1} a_n(x)=\rho_n(x)+\tilde{\rho_n}(x)+ \sum\limits_k
T_{jk}\ast\rho_k(x). \ee Here we introduced the notations
\begin{eqnarray}
\label{aa} a_n(x)=\frac 1\pi \;\frac n{x^2+n^2},
\end{eqnarray}
for $n=1\dots N$. The functions $\rho_n(x)$ and $\tr_n(x)$, which
are unknown, denote particle and hole densities, respectively. The
index $n$ represents their correspondence to $n$-strings. The
convolution operation, $"\ast"$, is defined as
\begin{eqnarray}
f\ast g\;(x)=\int_{-\infty}^{+\infty}f(x-y)g(y)dy,
\end{eqnarray}
for any given pair of functions $f$ and $g$. This is the
conventional definition. The functions $T_{nm}(x)$ for $n,\; m
=1\dots N$, involve the expressions $a_n(x)$ in their definition.
They have the following form

\be{T} T_{nm}(x)\equiv \left\{\matrix{a_{|n-m|}(x)+2a_{|n-m|+2}(x)
+2a_{|n-m|+4}(x)+...\qquad\qquad \qquad \cr \qquad \qquad\qquad
...+2a_{n+m-2}(x)+a_{n+m}(x)~~~~{\rm for}\quad n\ne m, \cr \cr
2a_{2}(x)+2a_{4}(x)+... +2a_{2n-2}(x)+a_{2n}(x)\quad {\rm
for}\quad n=m.}\right. \nonumber\\\ee

\vspace{3mm} As we have already mentioned, the equations
(\ref{dis1}) represent the thermodynamic limit of the Bethe
Equations. These are integral equations with respect to particle
and hole densities, containing all of the information about the
energy spectrum. Suppose that the system is in a state
characterized by densities $\rho_j(x)$ and $\tr_j(x)$. Then the
equilibrium dynamics of the system at temperature $T$ can be
extracted by minimizing the free energy,
 $F=E-TS$,
 with respect to the independent $\rho_j$.
 This yields the following non-linear integral
 equations for functions $\eta_n(x)=\tr_n(x)/\rho_n(x)$,

\be{GaudTaka} \ln\eta_n=\frac {g_n}T +
\sum\limits_{k=1}^{\infty}T_{nk} \ast\ln(1+\eta^{-1}_k),\qquad
g_n= -8\pi \left(1+2\alpha \partial_{x}\right)a_n. \ee

In order to analyze the ground state energy of the mean-field
model Eq.~(\ref{sh}) which is under our current consideration, we
need to go to the zero temperature limit in Eq.~(\ref{GaudTaka}).
For this purpose, let us introduce a set of new functions,
$\epsilon _n(x)$, as $\eta _n(x)=\exp\{\epsilon _n(x)/T\}$,
$n=1\dots N$, and substitute them into Eq.~(\ref{GaudTaka}). Then,
in the zero temperature limit, $T\rightarrow 0$,
Eq.~(\ref{GaudTaka}) acquires the following form

\begin{eqnarray}
\label{es}
&&\epsilon _1(x)=-8\pi \left(1+2\alpha \partial_{x}\right)s(x)+s\ast\epsilon_2^{\dagger}(x),\\
&&\epsilon_n(x)= s\ast(\epsilon _{n-1}+\epsilon
_{n+1})(x),\;\;\;\; n\geq 2,\nonumber
\end{eqnarray}
where the function $s(x)$ is defined as $s(x)=\large[4\cosh(\pi x/
2)\large]^{-1}$. The action of the dagger (minus), $\dagger$ (-),
in Eq.~(\ref{es}), leaves only the positive (negative) part of the
corresponding function, $\epsilon_n(x)$, as
\begin{eqnarray}
\label{epsilon} \epsilon^{\dagger}_n(x)=\left
\{{\begin{array}{ccc}
\epsilon_n(x) &\mbox{if}&\quad \epsilon_n(x)\geq 0\\
\,&\,&\,\\
0&\mbox{if}&\quad \epsilon_n(x)< 0,
\end{array} }\right .
\nonumber\\
\nonumber\\
\epsilon^-_n(x)=\epsilon_n(x)-\epsilon^{\dagger}_n(x).\qquad\qquad\qquad
\end{eqnarray}
By definition, all $\epsilon_n(x)$ are measured in units of $kT$
(where we set $k=1$), and therefore have magnitudes of energy.
Equations (\ref{es}) unambiguously define the solutions for
functions $\epsilon_n(x)$ provided that

\begin{eqnarray}
\label{mag_field}
\lim_{n\rightarrow\infty}\frac{\epsilon_n(x)}{n}=2B,
\end{eqnarray}
where $B$ is the "magnetic field", which in our case,
Eq.~(\ref{sh}), is zero. It is transparent from Eq.~(\ref{es}),
that $\epsilon _n(x)>0$ for $ n=2,3,..$, and only the function
$\epsilon _1(x)$ can be positive, as well as negative (can change
its sign crossing the $x$ axis at a certain point). The solution
of the system (\ref{es}) can be then expressed in terms of
$\epsilon _1(x)$ as

\bea{ses} &&\epsilon
_n(x)=\epsilon^{\dagger}_n(x)=a_{n-1}\ast\epsilon^{\dagger}_1(x)+2(n-1)B,
\quad n=2,3...\nonumber\\
&&\epsilon_1(x)=-8\pi \left(1+2\alpha \partial_{x}\right)s(x)+
\int_{\epsilon _1>0}(s\ast a_1)(x-y)\epsilon_1(y)dy.\nonumber \eea
From the last equation we see that if there exists such a point,
$x=a$, where the function $\epsilon _1(x)$ changes its sign (and,
therefore, $\epsilon _1(a)=0$), then $\pi \alpha >1$. Thus, one
arrives at a Wiener--Hopf type integral equation for the unknown
function $\epsilon _1(x)$
\begin{eqnarray}
\label{WH} \epsilon_1(x)=\epsilon _0(x)+\int\limits_{y\geq
a}R(x-y)\epsilon_1(y)dy,
\end{eqnarray}
where
\begin{eqnarray}
\label{ER} \epsilon _0(x)=-8\pi \left(1+2\alpha
\partial_{x}\right)s(x), \quad R(x)=\int\limits_{-\infty}^{\infty}\frac{d\omega}{2\pi}e^{-i\omega
x}\frac{e^{-|\omega |}}{2\cosh \omega }. \nonumber
\end{eqnarray}
The same kind of non-linear integral equation, occurring when the
so called disturbance term, $\epsilon _0(x)$, is not an even
function of $x$ and changes its sign, appears in the solutions of
staggered zigzag ladders with broken one-step translation symmetry
\cite{MS},~\cite{AST}. Therefore, according to our experience
drawn from the previous works, we assume that $\epsilon_1(x)=0$
for $x<a$. Then, Eq.~(\ref{WH}) will be valid for $x\geq a$. This
assumption does not affect the right hand side of Eq.~(\ref{WH})
and, without loss of generality, gives the same solution. This
solution can be obtained following the standard steps of the
technique of Wiener--Hopf integral equations. First, we apply a
Fourier transform
$$f(x)=\int\limits_{-\infty}^{\infty}\frac{d\omega}{2\pi}e^{-i\omega x}\tilde{f}(\omega),\quad
\tilde{f}(\omega) =\int\limits_{-\infty}^{\infty}dx e^{i\omega
x}f(x),
$$ to the functions $\epsilon_0(x)$, $\epsilon_1(x)$ and
$R(x)$. The substitution of these functions in the form of Fourier
integrals into Eq.~(\ref{WH}) yields \be{WHinteq}
\int\limits_{-\infty}^{\infty}\frac{d\omega}{2\pi}e^{-i\omega
x}\left\{[1-\tilde{R}(\omega)]\tilde{\epsilon}_1(\omega)-\tilde{\epsilon}_0(\omega)\right\}
=0,\quad x\geq a. \ee The equation (\ref{WHinteq}) can be
rewritten in an equivalent form, as \be{WHalgeq}
[1-\tilde{R}(\omega)]\tilde{\epsilon}_1(\omega)-\tilde{\epsilon}_0(\omega)=e^{i\omega
a}h_-(\omega), \ee where $h_{\pm}(\omega)$ are the boundary values
of analytic functions which do not have poles in the upper
($\Pi_+$) and lower ($\Pi_-$) complex half-planes respectively,
and have zero limiting values at corresponding infinite points.
Hence, with our assumption, we will have
$\tilde{\epsilon}_1(\omega)=e^{i\omega
a}\tilde{\epsilon}_+(\omega).$ The kernel in Eq.~(\ref{WHinteq})
can be factorized. It is precisely this factorization property of
the kernel which is responsible for the solvability of
Eq.~(\ref{WHinteq}). This means that the kernel can be represented
as a product \be{fac} [G_+(\omega)G_-(\omega)]^{-1}\equiv
1-\tilde{R}(\omega)=\frac{e^{|\omega|}}{2\cosh[\omega]}, \ee where
$G_{\pm}(\omega)$ are the boundary values of the analytic
functions which do not have zeroes or poles on $\Pi_{\pm}$,
respectively, and have the property $G_{+}(\infty)=G_-(\infty)=1$.
Then, for the Fourier components $\tilde{\epsilon}_1(\omega)$ in
Eq.~(\ref{WH}), the solution, when $x\geq a$, will be \be{swh}
\tilde{\epsilon}_+(\omega)=G_+(\omega)\mbox{\Large
P}_{+}\left[G_-(\omega)e^{-i\omega
a}\tilde{\epsilon}_0(\omega)\right]. \ee Here, the operators
$P_{\pm}$ are projectors, defined as $P_{\pm}[f(\omega
)]=f_{\pm}(\omega)$, for any given function $f(x)$. For example,
the action of the projector $P_{+}$ on the sum of the Fourier
components of the function $f(x)$ and a constant $c$, yields
\begin{eqnarray} \label{PP} \mbox{\Large
P}_{+}\left[c+\int_{-\infty}^{+\infty}\!dx\; e^{i \omega x}
f(x)\right]=c + \int_0^{+\infty}\!dx\;e^{i \omega x}f(x).
\end{eqnarray}
The complex functions $G_{+}(\omega )$ and $G_{-}(\omega )$, from
the factorization equation (\ref{fac}), can be calculated exactly.
They have the following algebraic forms \be{WHfactosr} G_-(\omega
)=\sqrt{2\pi }\;\;\frac{\exp\{-i\frac\omega \pi +i\frac\omega \pi
\ln\large( i\frac\omega \pi\large)\} }{\Gamma\left( \frac
12+i\frac\omega \pi \right)},\qquad G_+(\omega )=G_-(-\omega ).
\ee

Now one can derive the explicit form of the solution (\ref{swh})
for $a>0$. In order to do this, one just has to make use of the
above mentioned property of projectors $P_{+}$ and $P_{-}$, given
by Eq.~(\ref{PP}). Namely, upon application of Eq.~(\ref{PP}) to
the expression in brackets in the right hand side of
Eq.~(\ref{swh}), one will express the solution Eq.~(\ref{swh}) in
the form of an infinite sum
 \be{sswh}
\tilde{\epsilon}_+(\omega)=-i4\pi
G_+(\omega)\sum_{k=0}^{\infty}(-1)^k[1-2\pi \alpha
(k+1/2)]\frac{e^{-\pi a(k+1/2)}G_-(-i\pi (k+1/2))} {\omega +i\pi
(k+1/2)}. \ee The first two terms of the sum in this equation have
been obtained in Ref. \cite{tsvelik}. Now, in order to calculate
the free energy in the mean-field model Eq.~(\ref{sh}), we need to
find the parameter $a=a(\alpha )$, defined by the condition
$\epsilon_1(a)=0$. This condition can be rewritten as
\begin{eqnarray}
\label{zero} 0=\epsilon _1(a)=\int \frac {d\omega}{2\pi
}\tilde{\epsilon}_+(\omega)\simeq \frac i2 \lim _{|\omega
|\rightarrow \infty}\omega \tilde{\epsilon}_+(\omega).
\end{eqnarray}
Substituting the solution Eq.~(\ref{sswh}) for
$\tilde{\epsilon}_+(\omega)$ into Eq.~(\ref{zero}), we will
represent Eq.~(\ref{zero}) in an equivalent form as
 \be{a}
\sum_{k=0}^{\infty}(-1)^k[1-2\pi \alpha (k+1/2)]e^{-\pi
a(k+1/2)}G_-(-i\pi (k+1/2))=0. \ee

\begin{figure}[frst]
\begin{center}
\includegraphics[width=6.8cm,angle=0]{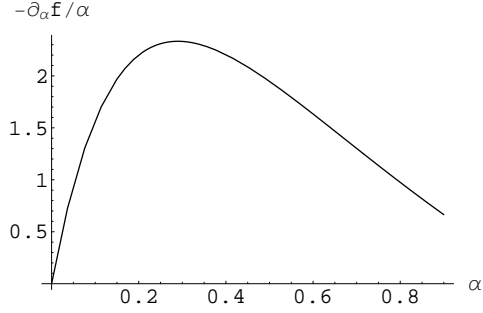}
\vspace{-0.2cm}
 \caption{The function $-{\partial_{\alpha}F(\alpha)}/\alpha$ versus $\alpha$} \label{fig1}
\end{center}
\end{figure}
\begin{figure}[frst]
\begin{center}
{$\includegraphics[width=6cm,angle=0]{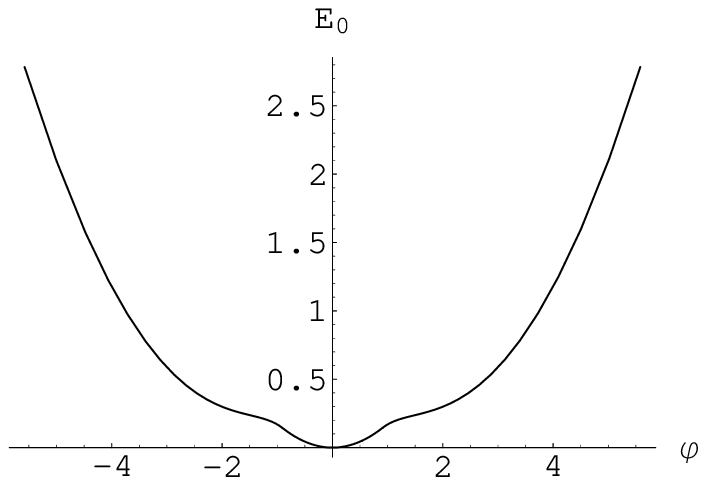}\atop {g=0.35}$}
{$\includegraphics[width=6cm,angle=0]{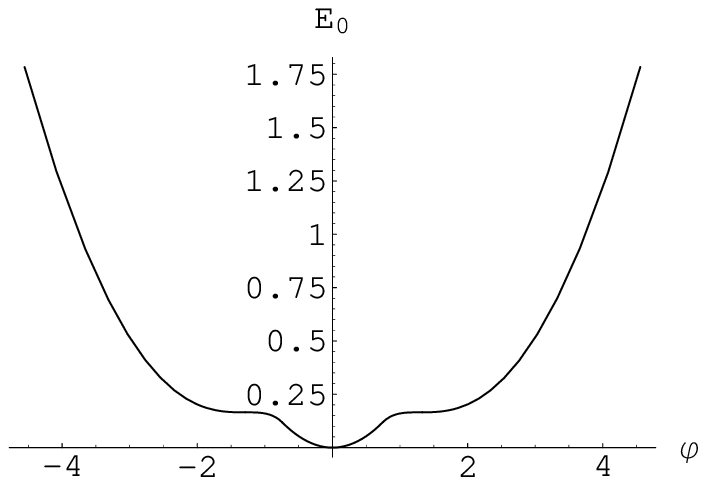}\atop {g=g_c}$}
{$\includegraphics[width=6cm,angle=0]{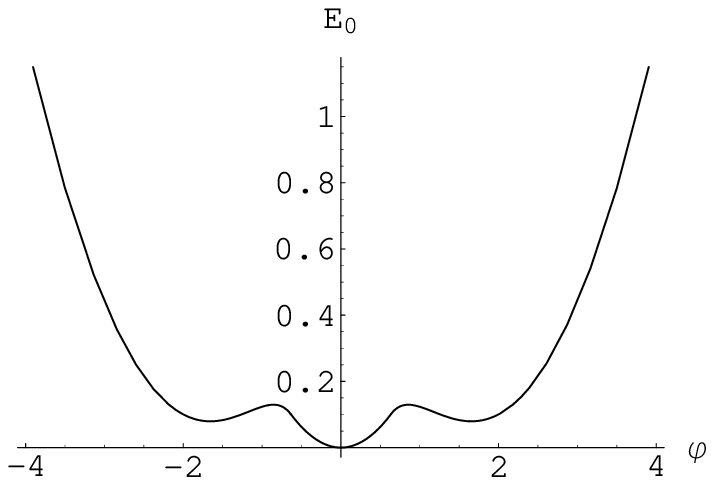}\atop {g=0.5}$}
{$\includegraphics[width=6cm,angle=0]{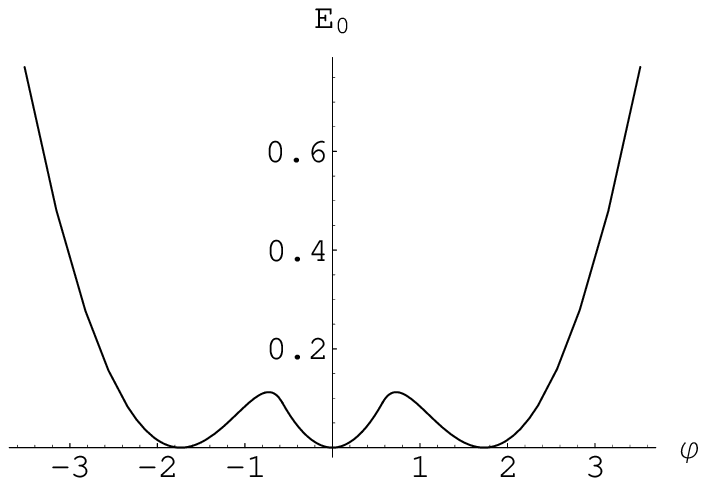}\atop
{g=g_{c2}}$}
{$\includegraphics[width=6cm,angle=0]{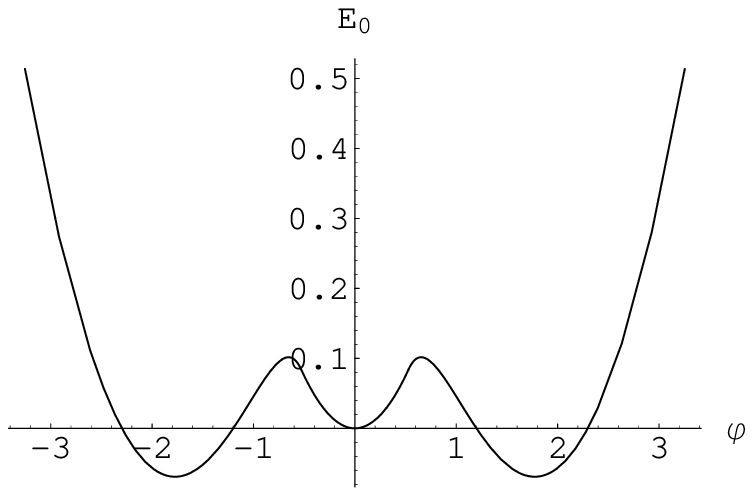}\atop {g=0.6}$}
 \caption{Vacuum energy $E_0$ versus parameter $\varphi$ for different values of $g$.} \label{5figs}
\end{center}
\end{figure}
The free energy per site at $T=0$, which is the ground state
energy, will explicitly depend on $a$. From the definition of the
free energy, we have
\begin{eqnarray}
\label{fe} F(\alpha )&=&-(4\log
2-1)-\int_{x\in L_{+}}\!\!s(x)\epsilon _1(x)dx\nonumber\\
&=&-(4\log
2-1)-\int\limits_{-\infty}^{\infty}\frac{d\omega}{2\pi}e^{-i\omega
a}\tilde{\epsilon}_+(\omega) s_+(-\omega ).
\end{eqnarray}
Here, the integration range of the first integral, $L_{+}$, is
given by those values of $x$, where $\epsilon _1(x)$ is positive.
Thus, substituting Eq.~(\ref{sswh}) into Eq.~(\ref{zero}), we get
the {\em exact} ground state energy of the model Eq.~(\ref{sh}) as
\begin{eqnarray}
\label{free} F(\alpha )=4-4\log 2
+2\sum_{n,k=0}^{\infty}(-1)^{k+n}[1-2\pi \alpha (k+1/2)]\qquad\qquad\qquad\\
\qquad\qquad\qquad\qquad\times \;\frac{e^{-\pi a(k+n+1)}G_+(i\pi
(n+1/2))G_-(-i\pi (k+1/2))} {k+n+1}.\nonumber
\end{eqnarray}

\section{Effective action and phase transitions}
\indent

The expression (\ref{free}), for the energy per site, can be
considered as an {\em effective action} for the order parameter
$\varphi$, defined by Eq.~(\ref{order}). The explicit form of the
effective action, Eq.~(\ref{free}), allows for further
investigations; in particular, with regard to the matter of
critical behavior, one can analyze in details the saddle point
equation corresponding to our mean-field theory. In terms of the
parameter $\alpha$, the saddle point equation (\ref{sp}) reads
\be{asp}
\partial_{\alpha}F(\alpha)=-\alpha /g,\quad\mbox{or,
if}\quad\alpha\neq 0,\quad
-\frac{\partial_{\alpha}F(\alpha)}\alpha=1/g. \ee For any given
$g$, this equation always has a zero solution $\alpha=0$. In order
to find a non-zero solution, we performed a numerical evaluation
of the function $-{\partial_{\alpha}F(\alpha)}/\alpha$ (where the
analytical form of $F(\alpha)$ is given by Eq~(\ref{free})), with
the condition given by Eq.~(\ref{a}). The plot is presented in
Fig.\ref{fig1}. From this picture, one can conclude that
Eq.~(\ref{asp}) has a solution when and only when $g$ exceeds the
critical value $g_c$, where $1/g_c$ equals to the maximal value of
$-{\partial_{\alpha}f(\alpha)}/\alpha$. For that value, our
calculations give $g_c=0.428646$. Then, from Eq.~(\ref{tj1j2}),
one can find the corresponding critical ratio
$$(J_2/J_1)_c=\frac{g_c}{1+2g_c}=0.230791,$$
which is in a good agreement with the expected value. It is also
interesting to investigate the behavior of the effective
potential, Eq.~(\ref{free}) versus the order parameter $\varphi$
for different values of $g$. The plots are presented in
Fig.\ref{5figs}. When $g$ is less than $g_c$, there is only one
vacuum energy minimum at $\varphi=0$ while, for $g\geq g_c$, two
new minima appear. There exists another value of $g$, which we
mark as $g_{c2}$
($g_{c2}=0.555083,\,\,\bigl[J_2/J_1\bigr]_{c2}=0.263052$), which
occur when the magnitudes of three vacuum energy minima are the
same. At this point we do not have an additional phase transition,
since the order parameter, $\varphi$, is smooth and finite.
However, it would be interesting to understand the reason and
consequences of such behavior.

Two non-zero minima become infinitely deeper and the positions of
minima approach to zero from both, left and right hand sides, as
$g$ further goes up to $+ \infty$. This scenario corresponds to
the Majumdar-Ghosh limit, $J_2/J_1 \rightarrow 1/2$, suggesting
the next phase transition where the chiral order parameter
$\varphi$ vanishes and another, fully dimerised phase appears.
Thus, our description of the intermediate phase with non-zero
chiral order parameter $\varphi$ complements the understanding of
the fully dimerised phase for $J_2/J_1\geq 1/2$.

\vspace{0.5cm} \noindent {\bf Concluding remarks.} We have derived
an effective action for the spin-1/2 $J_1-J_2$ model as a function
of the spin-liquid order parameter $\varphi=\langle\vec \sigma
_{n-1}(\vec \sigma _{n}\times \vec \sigma _{n+1})\rangle$, and
have observed the presence of two phase transitions at points
close to the expected values: {\em (i)} when $(J_2/J_1)<0.230791$,
we have an ordinary critical phase of isotropic  Heisenberg model;
{\em (ii)} in the middle phase, when $0.230791<(J_2/J_1)<1/2,$ the
ground state is $Z_2$ degenerate with two signs of order
parameter. Due to this degeneracy, kink-like topological
excitations are present and their condensation may characterize
the third phase at $J_2 / J_1
> 1/2$. Though the described picture do not coincide, but at the
same time is not in contradiction with the well known description
of this phase in the thermodynamic limit, when the two states, one
with wave vector $k=0$ (ground state for finite system) and
another with $k=\pi$ (first exited state for finite system),
collapse to each other and give rise to the dimerization pattern
(two-fold degeneracy) and the breaking of one-step translational
invariance.

In our opinion, the developed approach based on the chiral order
parameter $\varphi$, is alternative to the known methods for
description of the intermediate phase and provides promising
possibility to investigate this important problem further.

\section{Acknowledgment} The authors are indebted to A. G. Sedrakyan for
support and illuminating discussions. It is a pleasure to thank
A.A. Nersesyan and A.B. Zamolodchikov for very useful discussions.
V.M. acknowledges INTAS grants 03-51-5460, YSF 05-109-5041, and
Volkswagen Foundation of Germany for financial support.

\end{document}